\documentclass[
  aps,
  reprint,
  a4paper,
  superscriptaddress,
  nofootinbib
]{revtex4-2}

\usepackage[T1]{fontenc}
\usepackage[utf8]{inputenc}
\usepackage{amsmath,amssymb}
\usepackage{graphicx,xcolor}
\usepackage{soul}
\usepackage{microtype}
\usepackage[
  colorlinks=true,
  linkcolor=blue,
  citecolor=blue,
  urlcolor=blue
]{hyperref}

\begin{document}


\title{Continuous-wave 148.4-nm generation in top-seeded solution-grown strontium tetraborate}

\author{Yanzhang Wu}
\thanks{These authors contributed equally to this work.}
\affiliation{State Key Laboratory of Low-Dimensional Quantum Physics, Department of Physics, Tsinghua University, Beijing 100084, China}

\author{Peng Yang}
\thanks{These authors contributed equally to this work.}
\affiliation{
School of Mechanical and Materials Engineering,
North China University of Technology (NCUT),
Beijing 100144, China
}

\author{Lingfeng Yan}
\email[Corresponding author: ]{yanlingfeng136@gmail.com}
\affiliation{State Key Laboratory of Low-Dimensional Quantum Physics, Department of Physics, Tsinghua University, Beijing 100084, China}

\author{Qi Xiao}
\affiliation{State Key Laboratory of Low-Dimensional Quantum Physics, Department of Physics, Tsinghua University, Beijing 100084, China}

\author{Jiahong Li}
\affiliation{State Key Laboratory of Low-Dimensional Quantum Physics, Department of Physics, Tsinghua University, Beijing 100084, China}

\author{Juxian Li}
\affiliation{State Key Laboratory of Low-Dimensional Quantum Physics, Department of Physics, Tsinghua University, Beijing 100084, China}

\author{Beichen Huang}
\affiliation{State Key Laboratory of Low-Dimensional Quantum Physics, Department of Physics, Tsinghua University, Beijing 100084, China}

\author{Zhiwei Jiao}
\email[Corresponding author: ]{jzwzwj@163.com}
\affiliation{
School of Mechanical and Materials Engineering,
North China University of Technology (NCUT),
Beijing 100144, China
}

\author{Shiqian Ding}
\email[Corresponding author: ]{dingshq@mail.tsinghua.edu.cn}
\affiliation{State Key Laboratory of Low-Dimensional Quantum Physics, Department of Physics, Tsinghua University, Beijing 100084, China}
\affiliation{Beijing Academy of Quantum Information Sciences, Beijing 100193, China}
\affiliation{Frontier Science Center for Quantum Information, Beijing 100084, China}

\begin{abstract}
We report an implementation of continuous-wave (CW) vacuum-ultraviolet (VUV) generation at 148.4 nm by single-pass second-harmonic generation of 296.8 nm light in strontium tetraborate (SBO). At an incident fundamental power of approximately 580~mW, we estimate a VUV output power of approximately 1.7 nW at the crystal exit. The crystal is grown by the top-seeded solution growth method and operated in a helium-filled chamber. The quadratic power dependence observed under vacuum, together with its strong suppression in air, supports the identification of the generated second harmonic. We further characterize the dependence of the VUV signal on beam position, incidence angle, polarization, and crystal temperature, providing practical guidance for SBO-based CW VUV sources for nuclear spectroscopy and clock applications.
\end{abstract}

\maketitle

\section{Introduction}

The low-energy isomeric transition in $^{229}$Th provides
a unique opportunity to establish an optical frequency
reference based on a nuclear
transition~\cite{ClockPeik,CrystalHudson,ClockKuzmich,CrystalSchumm}.
Its transition energy of about 8.4~eV (about 148.4 nm), established through a series of
measurements of the isomer energy and radiative decay
~\cite{1976Energy,1990Energy,1994Energy,2007Energy,LarsDirect,2019EnergyMunich,2019EnergyJapan,
2020Energy,2022EnergyCERN}, makes it accessible to laser
spectroscopy, opening prospects for precision timekeeping
and sensitive tests of fundamental physics, including
searches for variations of fundamental constants and
ultralight dark
matter~\cite{ThoriumReviewNuClock,FlambaumVariation,
DarkMatter2025,Beeks2025NatCommunAlpha,DarkMatter2025,Beeks2025NatCommunAlpha}.

Direct laser excitation of the isomer and precision
spectroscopy with vacuum-ultraviolet (VUV) frequency combs
have established the experimental foundation for nuclear
clocks~\cite{LaserSpecSchumm,LaserSpecHudson,LaserSpecYe}.
In crystalline hosts, spectroscopy has further resolved
quadrupole structure, site-dependent shifts, temperature
sensitivities, and reproducible transition frequencies across
independently prepared crystals~\cite{LaserSpecYe,Hiraki2025LaserMossbauer,
Perera2025HostOffsets,Temperature,FrequencyReproducibility2025,Morawetz2026ContinuousCW,Huang2026NuclearClock}.
More recently, nuclear-clock operation has been demonstrated
by stabilizing continuous-wave (CW) VUV lasers to nuclear
resonances in $^{229}$Th-doped CaF$_2$
crystals~\cite{Huang2026NuclearClock,ToscaniDeCol2026NuclearClock}.
These advances motivate the development of practical
VUV sources for nuclear spectroscopy and clock interrogation.

Narrow-linewidth CW radiation near 148.4~nm enables sustained
interrogation of the nuclear resonance and feedback
stabilization of the probe frequency.
Several distinct approaches to CW
generation at this wavelength have now been demonstrated.
Resonance-enhanced four-wave mixing in cadmium vapour
provides narrow-linewidth VUV radiation with microwatt-level
output~\cite{Mat_elem,DingVUVproposal2024,DingVUV2025,Huang2026NuclearClock}.
An alternative route is nonlinear frequency conversion in
VUV-transparent crystals. CW generation near 148~nm has been
demonstrated by single-pass SHG using random quasi-phase matching
(RQPM) in strontium tetraborate
(SBO, SrB$_4$O$_7$)~\cite{PeikCWVUV}, by cavity-enhanced SHG in
periodically poled barium magnesium fluoride
(BMF, BaMgF$_4$)~\cite{Subramanian2026BMF}, and by engineered
high-$\chi^{(2)}$ quasi-phase-matched crystals~\cite{PerlovPatent2024, Vasilyev2026VUV}, with microwatt
CW output~\cite{YeICAP29_2026}.
These crystal-based schemes differ in nonlinear medium,
phase-matching strategy, optical implementation, and conversion
performance.

Among these approaches, RQPM in SBO offers an attractive
balance between output power and implementation complexity.
It exploits the irregular domain structure formed during
crystal growth, avoiding the need for a deliberately
patterned domain structure.
Early studies established the VUV transparency of
SBO~\cite{Pan2002SBO}, while pulsed SHG experiments
demonstrated ultraviolet conversion in Li-doped
SBO~\cite{Wu2024LiSBO} and VUV generation at wavelengths
down to 121~nm in randomly quasi-phase-matched
SBO~\cite{SBO2016}.
More recently, Lal \textit{et al.}\ demonstrated single-pass
CW SHG from 296.8~nm to 148.4~nm, obtaining an output power
of $1.3^{+0.7}_{-0.6}$~nW for an incident fundamental power
of 325~mW.
Their measurements also revealed a strong dependence
of the conversion performance on beam position,
incidence angle, and crystal temperature~\cite{PeikCWVUV}.
This sensitivity to the domain structure and operating
conditions motivates the characterization of independently
grown SBO crystals to assess the reproducibility and
practical utility of this approach.

Here, we report an independent implementation of CW SHG
at 148.4~nm using an SBO crystal grown by the top-seeded
solution growth method.
We describe the crystal preparation and structural
characterization, together with an experimental setup
comprising a helium-filled SHG chamber and a separate
filtered photon-counting detection chamber.
To identify the generated second harmonic, we measure
the fundamental-power dependence of the PMT count rate
with the detection chamber evacuated and filled with air.
The vacuum data exhibit a quadratic contribution,
whereas the signal remaining in air is approximately
linear in fundamental power.
We estimate the VUV output power
from the quadratic coefficient
of the vacuum-data fit, accounting for the optical
transmission and detector response.
We also characterize the dependence of the conversion
signal on beam position within the crystal, incidence
angle, fundamental polarization, and crystal temperature.
These measurements assess the practical performance
of an independently grown SBO crystal for CW VUV
generation.

\section{SBO crystal preparation and characterization}
\label{sec:crystal}

\begin{figure}[htbp]
    \centering
    \includegraphics[width=\linewidth]{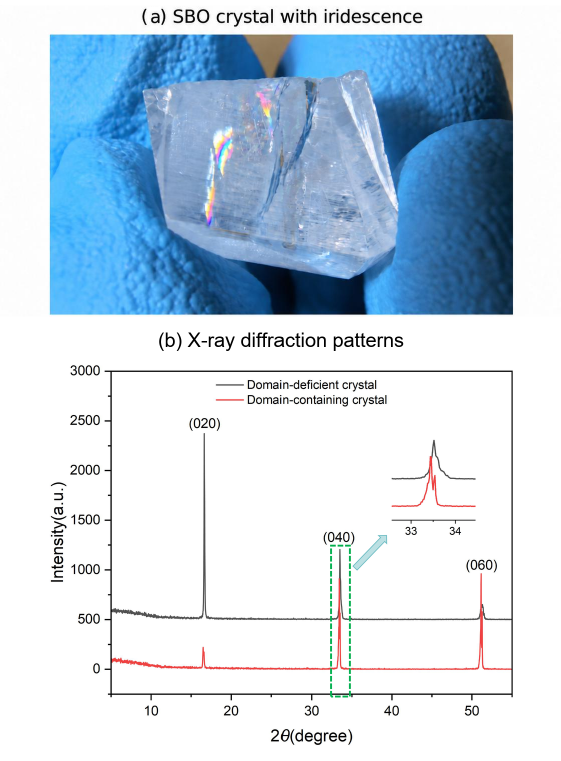}
    \caption{
    Characterization of the SBO crystal.
    (a) Representative photograph of an SBO crystal showing iridescence. 
    (b) X-ray diffraction patterns showing the $(0k0)$ reflections of the domain-containing and domain-deficient SBO samples.
    The inset shows an enlarged view of the $(040)$ reflection, highlighting the peak splitting.
    }
    \label{fig:sbo_xrd}
\end{figure}

The pure SBO crystal used in this study was grown by the
top-seeded solution growth (TSSG) method and exhibits
crystallization characteristics similar to those of lightly Li$^{+}$-doped crystals~\cite{Wu2024LiSBO}. 
During crystal growth, local perturbations, particularly temperature fluctuations, can induce wedge-shaped inclusions along the $c$-axis growth direction and promote the formation
of sheet-like nonlinear domains predominantly oriented perpendicular to the crystallographic $a$ axis.
Pronounced iridescence is observed around internal cracks in the as-grown crystal when viewed along the $a$ axis
[Fig.~\ref{fig:sbo_xrd}(a)], providing a qualitative indication
of fine-scale structural inhomogeneity in these regions.
This coloration may arise from domain-related microstructures
and fracture steps exposed or formed during crack propagation.

A sample was cleaved from the domain-containing region along
the $c$-axis growth direction for structural characterization
and frequency conversion, while a second sample was cleaved
from the domain-deficient region along the $a$-axis growth
direction for comparison.
Figure~\ref{fig:sbo_xrd}(b) compares the X-ray diffraction
patterns of the two samples for the $(0k0)$ reflections.
Both samples exhibit the same principal reflections, indicating
the same underlying crystal phase.
The domain-containing sample, however, shows a shift toward
lower diffraction angles together with clear peak splitting.
These features are consistent with local strain and structural
variations associated with the domain-containing region.
In particular, local tensile strain near domain walls could
increase the interplanar spacing relative to that in the domain
interior. According to Bragg's law,
$2d\sin\theta=n\lambda$, an increase in $d$ shifts the
corresponding diffraction peak toward lower angle.
Variations in lattice spacing or crystallographic orientation
among different regions may also contribute to the observed
peak splitting, consistent with diffraction-profile changes
reported in materials containing domain
boundaries~\cite{Whatmore2025OPB,Cheary1991Antiphase}. 

\section{VUV generation and detection}
\label{sec:setup}

\begin{figure}[t]
    \centering
    \includegraphics[width=\columnwidth]{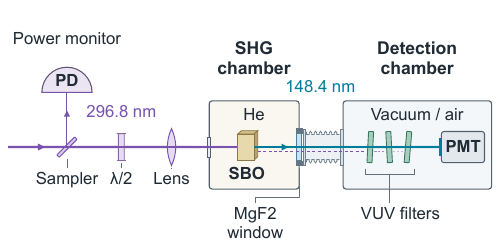}
    \caption{
    Experimental setup for CW SHG at 148.4~nm in SBO.
    A sampled fraction of the 296.8-nm fundamental beam is monitored
    by a photodiode (PD), while the main beam passes through a
    half-wave plate ($\lambda/2$) for polarization adjustment and is focused into the SBO crystal
    in a helium-filled SHG chamber.
    The generated VUV radiation exits through an MgF$_2$ window and
    is detected by a photomultiplier tube (PMT) after three VUV
    bandpass filters in a separate chamber operated under vacuum or air.
    Purple and cyan lines denote the fundamental and VUV beams,
    respectively.
    }
    \label{fig:setup}
\end{figure}

The SBO crystal used in this experiment has dimensions of
$2.5 \times 4.6 \times 7.6~\mathrm{mm}^{3}$ along the
crystallographic $a$, $b$, and $c$ axes, respectively. It is
mounted in a temperature-controlled holder that allows
adjustment of the transverse position and incidence
angle. Up to 580~mW of 296.8-nm fundamental light is
focused into the crystal, propagating approximately along the
$a$ axis with its polarization nominally aligned with the $c$ axis.
The beam has a $1/e^{2}$ intensity waist radius of approximately
$25~\mu\mathrm{m}$ at the crystal center~(Fig.~\ref{fig:setup}).

The crystal is housed in an SHG chamber filled with helium at
approximately $1\times10^{5}~\mathrm{Pa}$. The helium atmosphere
provides a VUV-transparent environment and facilitates heat
transfer from the crystal~\cite{PeikCWVUV}. The generated
radiation exits through an MgF$_2$ window and enters a separate
detection chamber.

During VUV measurements, the detection chamber is evacuated to
$5\times10^{-5}~\mathrm{Pa}$. As a control, the detection
chamber is filled with air, which strongly attenuates 148.4-nm
radiation. The resulting suppression of the detected signal
therefore distinguishes the VUV signal from background due to
residual fundamental light.

Three VUV bandpass filters are placed before a solar-blind CsI PMT
(Hamamatsu R10454) to suppress residual 296.8-nm light.
Each filter has a manufacturer-specified transmission of
approximately 21.4\% at 148.4~nm, whereas its measured
transmission at 296.8~nm is approximately $4\times10^{-4}$.
The filters are slightly tilted from normal
incidence to reduce parasitic interference and back-reflections.
The PMT is operated in photon-counting mode, with an estimated
quantum efficiency of approximately 22.2\% at 148.4~nm.
Based on the filter transmissions and the manufacturer's typical PMT spectral responsivity, their combined response provides approximately twelve orders of magnitude of nominal rejection of 296.8~nm light relative to 148.4~nm light for equal incident optical powers.

\section{Identification of the VUV signal and estimation of the output power}
\label{sec:conversion}
\begin{figure}[t]
    \centering
    \includegraphics[width=\columnwidth]{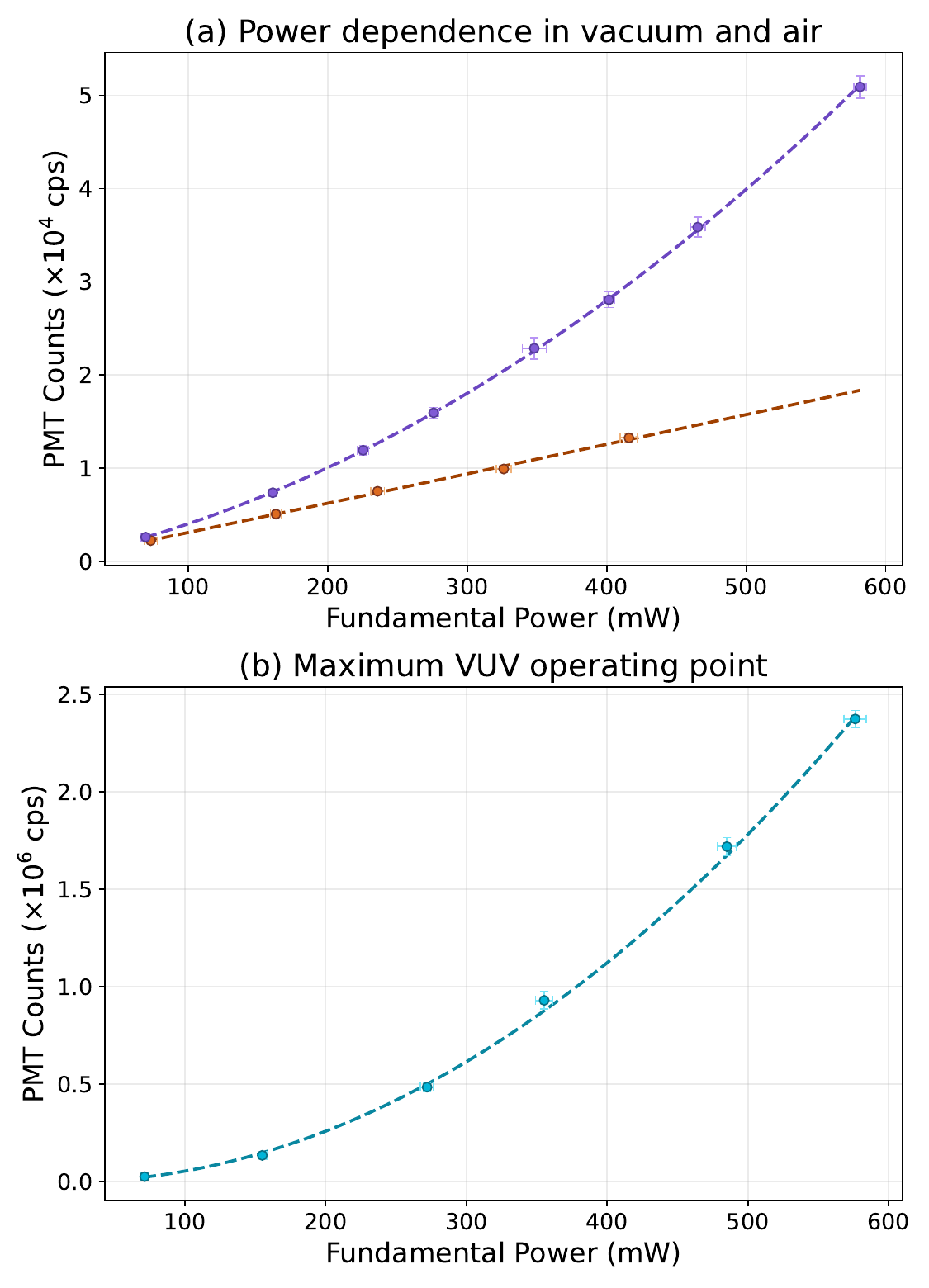}
    \caption{PMT count rate as a function of the fundamental power incident on the SBO crystal.
    (a) Initial identification of the generated VUV signal. Purple and orange points show measurements with the detection chamber evacuated and filled with air, respectively.
    Dashed curves show a quadratic-polynomial fit to the vacuum data [Eq.~\eqref{eq:vacuum_fit}] and a linear fit to the air data.
    (b) Measured PMT count rate as a function of the fundamental power at the optimized operating point. The data points represent the measured PMT count rates and the dashed curve shows a quadratic fit to the experimental data. Error bars in both plots indicate standard deviations.
}
    \label{fig:power}
\end{figure}

We first identify the generated VUV signal by measuring the
PMT count rate as a function of the fundamental power incident
on the crystal, $P_{\mathrm{UV}}$, with the detection chamber
either evacuated or filled with air.
As shown in Fig.~\ref{fig:power} (a), the vacuum data exhibit a
quadratic contribution, whereas admitting air substantially
reduces the count rate and leaves an approximately linear
power dependence.
Because air strongly attenuates light at 148.4~nm while
transmitting the residual 296.8~nm fundamental light over
the detection path, these observations support the assignment
of the quadratic contribution to VUV SHG.

\begin{figure*}[t]
    \centering
    \includegraphics[width=\textwidth]{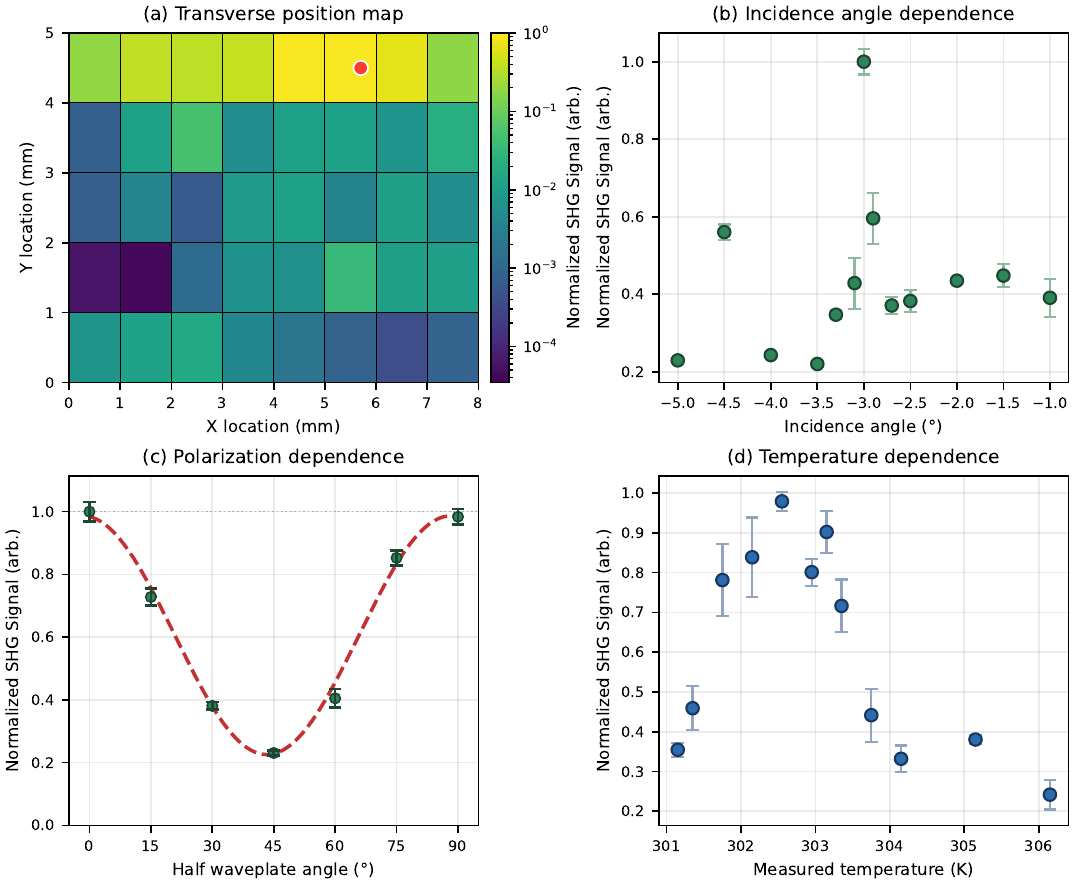}
    \caption{
    Spatial and parametric dependence of the SHG signal.
    (a) Transverse position map obtained by translating the
    SBO crystal perpendicular to the incident fundamental beam.
    The color scale is logarithmic. The red circle marks the
    position $(x,y)=(5.7,4.5)$~mm used for the scans in (b)--(d).
    (b) Dependence on the incidence angle.
    (c) Dependence on the half-wave plate angle controlling the
    polarization direction of the incident fundamental beam.
    The dashed curve represents a fit to the measured data.
    (d) Dependence on the measured SBO crystal temperature.
    Error bars in (b)--(d) indicate standard deviations.
    }
    \label{fig:scans}
\end{figure*}

The vacuum count rate is fitted over the full scanned
fundamental-power range using
\begin{equation}
C_{\mathrm{vac}}(P_{\mathrm{UV}})
= aP_{\mathrm{UV}}^2
+ bP_{\mathrm{UV}} + c,
\label{eq:vacuum_fit}
\end{equation}
where the quadratic term is attributed to SHG, the linear
term describes residual pump-related background, and the
constant term accounts for dark counts and power-independent
background.

{\color{black}{
The VUV power at the crystal exit is inferred from the fitted quadratic contribution
according to
\begin{equation}
P_{148}
= \frac{hc}{\lambda_{148}}\,
  \frac{aP_{\mathrm{UV}}^2}{\epsilon_{\mathrm{det}}},
\label{eq:vuv_power}
\end{equation}
where $\lambda_{148}=148.4$~nm and
$\epsilon_{\mathrm{det}}$ is the overall detection efficiency
from a VUV photon emerging from the crystal to a registered
count. It is given by
\begin{equation}
\epsilon_{\mathrm{det}}
= Q_{148}\,
  \epsilon_{\mathrm{pc}}\,
  \epsilon_{\mathrm{coll}}\,
  T_{\mathrm{w}}
  \prod_{i=1}^{3} T_i,
\label{eq:detection_efficiency}
\end{equation}
where $Q_{148}$ is the PMT quantum efficiency,
$\epsilon_{\mathrm{pc}}$ accounts for photoelectron collection
and pulse-discrimination acceptance,
$\epsilon_{\mathrm{coll}}$ is the optical collection efficiency,
and $T_{\mathrm{w}}$ and $T_i$ are the transmissions of the
MgF$_2$ window and the individual VUV filters, respectively.

For a nominal power estimate, we use
$Q_{148}\simeq0.222$, $T_{\mathrm{w}}\simeq0.90$, and
$T_i\simeq0.214$, and take
$\epsilon_{\mathrm{pc}}=\epsilon_{\mathrm{coll}}=1$.
These values give
$\epsilon_{\mathrm{det}}\simeq1.96\times10^{-3}$,
corresponding to approximately
$6.8\times10^{-16}~\mathrm{W/cps}$.

At the optimized operating point, the measured
fundamental-power dependence is shown in
Fig.~\ref{fig:power}(b).
The fit yields
$a=7.5~\mathrm{cps}/\mathrm{mW}^{2}$.
At an incident fundamental power of 580~mW, the corresponding
quadratic contribution is approximately
$2.5\times10^{6}~\mathrm{cps}$, giving an estimated VUV
output power of 1.7~nW at the crystal exit.

}}

\section{Spatial and parametric dependence of the SHG signal}

To identify operating conditions that maximize the SHG
signal, we systematically scan the fundamental beam
position and incidence angle at the crystal, the incident
polarization, and the SBO crystal temperature.
Figure~\ref{fig:scans} summarizes the measured dependence
of the normalized SHG signal on these parameters.

The transverse position is scanned by translating the
SBO crystal perpendicular to the incident beam using
the stage supporting the crystal mount.
The resulting map in Fig.~\ref{fig:scans}(a) reveals
strong spatial inhomogeneity, with the signal varying
by several orders of magnitude across the mapped region.
The strongest signals occur near the upper edge of the map.
The position marked by the red circle,
$(x,y)=(5.7,4.5)$~mm, is selected for the subsequent
parameter scans.

At this position, the incidence angle is varied by
adjusting the tilt of the crystal mount on the same stage.
As shown in Fig.~\ref{fig:scans}(b), the SHG signal
exhibits a nonmonotonic angular dependence, with a
pronounced maximum near $-3^\circ$ within the scanned
range.
The position and angle scans demonstrate the importance
of both selecting a favorable interaction region and
adjusting the crystal orientation.

We next investigate the polarization dependence using
the linearly polarized fundamental beam.
As described in Sec.~\ref{sec:setup}, a half-wave plate upstream of the SHG chamber is rotated to vary the polarization direction incident on the crystal.
Figure~\ref{fig:scans}(c) shows the measured signal as a function of the waveplate angle.
The signal is largest near waveplate angles of $0^\circ$ and $90^\circ$, corresponding to fundamental polarization along the crystal $c$ axis, and decreases to approximately one quarter of its maximum near $45^\circ$, where the polarization is perpendicular to the $c$ axis.
The corresponding rotation of the incident linear polarization is twice the waveplate rotation.

Finally, Fig.~\ref{fig:scans}(d) shows the dependence
of the SHG signal on the measured SBO crystal temperature.
The strongest signals are observed around 302--303~K,
with the largest measured signal near 302.6~K.
Together, these measurements identify favorable operating conditions for CW VUV generation in the present SBO crystal.

\section{Conclusion}

We have demonstrated single-pass continuous-wave second-harmonic generation from 296.8~nm to 148.4~nm in an independently grown SBO crystal prepared by the top-seeded solution growth method.
The crystal is operated in a helium-filled chamber, with the generated light detected in a separate evacuated photon-counting chamber.
A quadratic contribution to the count rate under vacuum, together with its suppression upon admitting air, supports the identification of VUV SHG.
From the fitted quadratic contribution and the estimated detection efficiency, we infer an output power of approximately 1.7~nW at the crystal exit for an incident fundamental power of 580~mW.

Spatial and parametric scans reveal strong variations across the crystal and identify favorable settings of incidence angle, polarization, and temperature.
The largest polarization-dependent signal is obtained with the fundamental polarization along the crystal $c$ axis.
These results demonstrate CW VUV generation in an independently grown SBO sample and provide practical guidance for developing crystal-based sources for $^{229}$Th nuclear spectroscopy and clock applications.

\par
\vspace*{12pt}
{\centering\small\bfseries FUNDING AND ACKNOWLEDGMENTS\par}
\nobreak\vspace{8pt}
This work is supported by Beijing Science and Technology Planning Project (No. Z25110100040000), the National Natural Science Foundation of China (Nos. 12341401 and 12274253), and Tsinghua University Dushi Program.



\bibliographystyle{unsrtnat}
\bibliography{Th}

@article{1976Energy,
  title={Features of the low-energy level scheme of $^{229}${T}h as observed in the $\alpha$-decay of $^{233}${U}},
  author={Kroger, LA and Reich, CW},
  journal={Nucl. Phys. A},
  volume={259},
  number={1},
  pages={29--60},
  year={1976},
  publisher={Elsevier}
}

@article{1990Energy,
  title={Energy separation of the doublet of intrinsic states at the ground state of $^{229}${T}h},
  author={Reich, CW and Helmer, RG},
  journal={Phys. Rev. Lett.},
  volume={64},
  number={3},
  pages={271},
  year={1990},
  publisher={APS}
}

@article{1994Energy,
  title={An excited state of $^{229}${T}h at 3.5 e{V}},
  author={Helmer, RG and Reich, CW},
  journal={Phys. Rev. C},
  volume={49},
  number={4},
  pages={1845},
  year={1994},
  publisher={APS}
}

@article{2007Energy,
  title={Energy splitting of the ground-state doublet in the nucleus $^{229}${T}h},
  author={Beck, BR and Becker, JA and Beiersdorfer, P and Brown, GV and Moody, Kenton J and Wilhelmy, Jerry B and Porter, FS and Kilbourne, CA and Kelley, RL},
  journal={Phys. Rev. Lett.},
  volume={98},
  number={14},
  pages={142501},
  year={2007},
  publisher={APS}
}

@article{2019EnergyMunich,
  title={Energy of the $^{229}${T}h nuclear clock transition},
  author={Seiferle, Benedict and von der Wense, Lars and Bilous, Pavlo V and Amersdorffer, Ines and Lemell, Christoph and Libisch, Florian and Stellmer, Simon and Schumm, Thorsten and D{\"u}llmann, Christoph E and P{\'a}lffy, Adriana and others},
  journal={Nature},
  volume={573},
  number={7773},
  pages={243--246},
  year={2019},
  publisher={Nature Publishing Group UK London}
}

@article{2019EnergyJapan,
  title={X-ray pumping of the $^{229}${T}h nuclear clock isomer},
  author={Masuda, Takahiko and Yoshimi, Akihiro and Fujieda, Akira and Fujimoto, Hiroyuki and Haba, Hiromitsu and Hara, Hideaki and Hiraki, Takahiro and Kaino, Hiroyuki and Kasamatsu, Yoshitaka and Kitao, Shinji and others},
  journal={Nature},
  volume={573},
  number={7773},
  pages={238--242},
  year={2019},
  publisher={Nature Publishing Group UK London}
}

@article{2020Energy,
  title={Measurement of the $^{229}${T}h isomer energy with a magnetic microcalorimeter},
  author={Sikorsky, Tomas and Geist, Jeschua and Hengstler, Daniel and Kempf, Sebastian and Gastaldo, Loredana and Enss, Christian and Mokry, Christoph and Runke, J{\"o}rg and D{\"u}llmann, Christoph E and Wobrauschek, Peter and others},
  journal={Phys. Rev. Lett.},
  volume={125},
  number={14},
  pages={142503},
  year={2020},
  publisher={APS}
}

@article{2022EnergyCERN,
  title={Observation of the radiative decay of the $^{229}${T}h nuclear clock isomer},
  author={Kraemer, Sandro and Moens, Janni and Athanasakis-Kaklamanakis, Michail and Bara, Silvia and Beeks, Kjeld and Chhetri, Premaditya and Chrysalidis, Katerina and Claessens, Arno and Cocolios, Thomas E and Correia, Jo{\~a}o GM and others},
  journal={Nature},
  volume={617},
  number={7962},
  pages={706--710},
  year={2023},
  publisher={Nature Publishing Group UK London}
}

@article{ThoriumReviewNuClock,
  title={Nuclear clocks for testing fundamental physics},
  author={Peik, Ekkehard and Schumm, Thorsten and Safronova, MS and Palffy, Adriana and Weitenberg, Johannes and Thirolf, Peter G},
  journal={Quantum Sci. Technol.},
  volume={6},
  number={3},
  pages={034002},
  year={2021},
  publisher={IOP Publishing}
}

@article{LarsDirect,
  title={Direct detection of the $^{229}${T}h nuclear clock transition},
  author={v. d. Wense, L. and Seiferle, B. and Laatiaoui, M. and Neumayr, J. B. and Maier, H. and Wirth, H. and Mokry, C. and Runke, J. and Eberhardt, K. and D{\"u}llmann, C. E. and Trautmann, N. G. and Thirolf, P. G.},
  journal={Nature},
  volume={533},
  number={7601},
  pages={47},
  year={2016},
  publisher={Nature Publishing Group}
}

@article{ClockPeik,
  title={Nuclear laser spectroscopy of the 3.5 e{V} transition in {T}h-229},
  author={Peik, E. and Tamm, C.},
  journal={Europhys. Lett.},
  volume={61},
  number={2},
  pages={181},
  year={2003},
  publisher={IOP Publishing}
}

@article{ClockKuzmich,
  title={Single-ion nuclear clock for metrology at the 19th decimal place},
  author={Campbell, C. J. and Radnaev, A. G. and Kuzmich, A. and Dzuba, V. A. and Flambaum, V. V. and Derevianko, A.},
  journal={Phys. Rev. Lett.},
  volume={108},
  number={12},
  pages={120802},
  year={2012},
  publisher={APS}
}

@article{FlambaumVariation,
  title={Enhanced effect of temporal variation of the fine structure constant and the strong interaction in $^{229}${T}h},
  author={Flambaum, V. V.},
  journal={Phys. Rev. Lett.},
  volume={97},
  number={9},
  pages={092502},
  year={2006},
  publisher={APS}
}

@article{CrystalHudson,
  title={Constraining the evolution of the fundamental constants with a solid-state optical frequency reference based on the $^{229}${T}h nucleus},
  author={Rellergert, Wade G and DeMille, D and Greco, Richard R and Hehlen, Markus Peter and Torgerson, JR and Hudson, Eric R},
  journal={Phys. Rev. Lett.},
  volume={104},
  number={20},
  pages={200802},
  year={2010},
  publisher={APS}
}

@article{CrystalSchumm,
  title={Performance of a $^{229}${T}horium solid-state nuclear clock},
  author={Kazakov, GA and Litvinov, AN and Romanenko, VI and Yatsenko, LP and Romanenko, AV and Schreitl, Matthias and Winkler, Georg and Schumm, Thorsten},
  journal={New J. Phys.},
  volume={14},
  number={8},
  pages={083019},
  year={2012},
  publisher={IOP Publishing}
}

@article{DarkMatter2025,
  title = {Searching for Dark Matter with the $^{229}\mathrm{Th}$ Nuclear Lineshape from Laser Spectroscopy},
  author = {Fuchs, Elina and Kirk, Fiona and Madge, Eric and Paranjape, Chaitanya and Peik, Ekkehard and Perez, Gilad and Ratzinger, Wolfram and Tiedau, Johannes},
  journal = {Phys. Rev. X},
  volume = {15},
  issue = {2},
  pages = {021055},
  numpages = {15},
  year = {2025},
  month = {May},
  publisher = {American Physical Society},
  doi = {10.1103/PhysRevX.15.021055},
}

@article{LaserSpecSchumm,
  title={Laser excitation of the {T}h-229 nucleus},
  author={Tiedau, J and Okhapkin, MV and Zhang, K and Thielking, J and Zitzer, G and Peik, E and Schaden, F and Pronebner, T and Morawetz, I and Toscani De Col, L and others},
  journal={Phys. Rev. Lett.},
  volume={132},
  number={18},
  pages={182501},
  year={2024},
  publisher={APS}
}

@article{LaserSpecHudson,
  title={Laser excitation of the $^{229}${T}h nuclear isomeric transition in a solid-state host},
  author={Elwell, R and Schneider, Christian and Jeet, Justin and Terhune, JES and Morgan, HWT and Alexandrova, AN and Tan, HB and Derevianko, Andrei and Hudson, Eric R},
  journal={Phys. Rev. Lett.},
  volume={133},
  pages={013201},
  year={2024},
  publisher={APS}
}

@article{SBO2016,
  title={Generation of coherent radiation in the vacuum ultraviolet using randomly quasi-phase-matched strontium tetraborate},
  author={Trabs, Peter and Noack, Frank and Aleksandrovsky, Aleksandr S and Zaitsev, Alexandre I and Petrov, Valentin},
  journal={Opt. Lett.},
  volume={41},
  number={3},
  pages={618--621},
  year={2016},
  publisher={Optical Society of America}
}

@article{LaserSpecYe,
  title={Frequency ratio of the $^{229\text{m}}${T}h nuclear isomeric transition and the $^{87}${S}r atomic clock},
  author={Zhang, Chuankun and Ooi, Tian and Higgins, Jacob S and Doyle, Jack F and von der Wense, Lars and Beeks, Kjeld and Leitner, Adrian and Kazakov, Georgy A and Li, Peng and Thirolf, Peter G and others},
  journal={Nature},
  volume={633},
  number={8028},
  pages={63--70},
  year={2024},
  publisher={Nature Publishing Group UK London}
}

@article{Temperature,
  title={Temperature sensitivity of a {T}horium-229 solid-state nuclear clock},
  author={Higgins, Jacob S and Ooi, Tian and Doyle, Jack F and Zhang, Chuankun and Ye, Jun and Beeks, Kjeld and Sikorsky, Tomas and Schumm, Thorsten},
  journal={Phys. Rev. Lett.},
  volume={134},
  number={11},
  pages={113801},
  year={2025},
  publisher={APS}
}

@article{DingVUV2025,
  title = {Continuous-wave narrow-linewidth vacuum ultraviolet laser source},
  author = {Xiao, Qi and Penyazkov, Gleb and Li, Xiangliang and Huang, Beichen and Bu, Wenhao and Shi, Juanlang and Shi, Haoyu and Liao, Tangyin and Yan, Gaowei and Tian, Haochen and Li, Yixuan and Li, Jiatong and Lu, Bingkun and You, Li and Lin, Yige and Mo, Yuxiang and Ding, Shiqian},
  journal = {Nature},
  volume = {650},
  pages = {852--856},
  year = {2026},
}

@article{DingVUVproposal2024,
  title = {Proposal for the generation of continuous-wave vacuum-ultraviolet laser light for {Th}-229 isomer precision spectroscopy},
  author = {Xiao, Qi and Penyazkov, Gleb and Yu, Ruihan and Huang, Beichen and Li, Jiatong and Shi, Juanlang and Yu, Yanmei and Mo, Yuxiang and Ding, Shiqian},
  journal = {Phys. Rev. Appl.},
  volume = {25},
  number = {2},
  pages = {024034},
  year = {2026},
}

@article{FrequencyReproducibility2025,
  author  = {Ooi, Tian and Doyle, Jack F. and Zhang, Chuankun and Higgins, Jacob S. and Ye, Jun and Beeks, Kjeld and Sikorsky, Tomas and Schumm, Thorsten},
  title   = {Frequency reproducibility of solid-state thorium-229 nuclear clocks},
  journal = {Nature},
  volume  = {650},
  pages   = {72--78},
  year    = {2026},
}

@article{Hiraki2025LaserMossbauer,
  title={Laser {M{\"o}ssbauer} spectroscopy of {$^{229}${T}h} in {CaF}$_2$},
  author={Hiraki, Takahiro and Masuda, Takahiko and Takatori, Sayuri and Schaden, Fabian and Bartokos, Michael and Beeks, Kjeld and Fukunaga, Yuta and Gr{\"u}neis, Andreas and Guan, Ming and Kazakov, Georgy and others},
  journal={Science},
  volume={393},
  number={6813},
  pages={795-799},
  year={2026},
  publisher={American Association for the Advancement of Science}
}

@article{Morawetz2026ContinuousCW,
  author    = {Morawetz, I. and Riebner, T. and Toscani De Col, L. and Schneider, F. and Sempelmann, N. and Schaden, F. and Bartokos, M. and Kazakov, G. A. and Lahs, S. and Beeks, K. and Gerstenecker, B. and Gr{\"u}neis, A. and Lal, V. and Zitzer, G. and Tiedau, J. and Okhapkin, M. V. and Petrov, V. and Peik, E.},
  title     = {Continuous-wave laser absorption spectroscopy of the {T}horium-229 nucleus},
  journal   = {Nature},
  volume = {657},
  pages = {626--631},
  year      = {2026},
  doi = {https://doi.org/10.1038/s41586-026-11011-7}
}

@article{PeikCWVUV,
author = {Vishal Lal and Maksim V. Okhapkin and Johannes Tiedau and Niels Irwin and Valentin Petrov and Ekkehard Peik},
journal = {Optica},
number = {12},
pages = {1971--1974},
publisher = {Optica Publishing Group},
title = {Continuous-wave laser source at the 148~nm nuclear transition of {T}h-229},
volume = {12},
month = {Dec},
year = {2025},
}

@article{Mat_elem,
  title = {Theoretical study of transition matrix elements in cadmium for vacuum-ultraviolet generation in $^{229}\mathrm{Th}$ nuclear clock applications},
  author = {Penyazkov, Gleb and Yu, Yanmei and Skripnikov, Leonid V. and Ding, Shiqian},
  journal = {Phys. Rev. A},
  volume = {112},
  issue = {2},
  pages = {022807},
  numpages = {8},
  year = {2025},
  month = {Aug},
  publisher = {American Physical Society}
}

@article{Beeks2025NatCommunAlpha,
  title   = {Fine-structure constant sensitivity of the {T}h-229 nuclear clock transition},
  author  = {Beeks, Kjeld and Kazakov, Georgy A. and Schaden, Fabian and Morawetz, Ira and Toscani de Col, Luca and Riebner, Thomas and Bartokos, Michael and Sikorsky, Tomas and Schumm, Thorsten and Zhang, Chuankun and Ooi, Tian and Higgins, Jacob S. and Doyle, Jack F. and Ye, Jun and Safronova, Marianna S.},
  journal = {Nat. Commun.},
  volume  = {16},
  pages   = {9147},
  year    = {2025},
  doi     = {10.1038/s41467-025-64191-7}
}

@article{Perera2025HostOffsets,
  title   = {Host-Dependent Frequency Offsets in {$^{229}${T}h} Nuclear Clockwork},
  author  = {Perera, U. C. and Morgan, H. W. T. and Hudson, Eric R. and Derevianko, Andrei},
  journal = {Phys. Rev. Lett.},
  volume  = {135},
  pages   = {123001},
  year    = {2025},
  doi     = {10.1103/mhwc-4m14}
}

@article{Wu2024LiSBO,
  author  = {Wu, Yuhao and Zhang, Fengfeng and Zhang, Shenjin
             and Zhao, Ying and Jiao, Zhiwei and Hou, Zhengzheng},
  title   = {Growth characteristics and {SHG} properties of
             {Li:SrB$_4$O$_7$}},
  journal = {Optical Materials},
  volume  = {152},
  pages   = {115433},
  year    = {2024},
  doi     = {10.1016/j.optmat.2024.115433}
}

@article{Whatmore2025OPB,
  author  = {Whatmore, Roger W. and Dutta, Debismita
             and Keeney, Lynette},
  title   = {A model for out-of-phase boundary induced
             {X-ray} diffraction peak profile changes
             in {Aurivillius} oxide thin films},
  journal = {Journal of Applied Crystallography},
  volume  = {58},
  number  = {4},
  pages   = {1191--1204},
  year    = {2025},
  doi     = {10.1107/S1600576725004091}
}

@article{Cheary1991Antiphase,
  author  = {Cheary, R. W. and Squadrito, R.},
  title   = {Electron and {X-ray} Diffraction from Antiphase
             Domains in the Barium Magnesium Hollandite
             {Ba$_{1.33}$Mg$_{1.33}$Ti$_{6.67}$O$_{16}$}},
  journal = {Acta Crystallographica Section B: Structural Science},
  volume  = {47},
  pages   = {1001--1008},
  year    = {1991}
}

@article{Huang2026NuclearClock,
  author  = {Huang, Beichen and Yan, Gaowei and Xiao, Qi
             and Bu, Wenhao and Zhao, Chengchun and Zhang, Zhen
             and Yan, Chao and others},
  title   = {A nuclear clock synchronized to {$^{229}$Th}},
  journal = {Nature},
  year    = {2026},
  note    = {In press}
}

@misc{ToscaniDeCol2026NuclearClock,
  author        = {{Toscani De Col}, L. and Riebner, T.
                   and Morawetz, I. and Schneider, F.
                   and Sempelmann, N.
                   and Schlachet-L{\'e}pinay, J. and others},
  title         = {A thorium-229 optical nuclear clock
                   with feedback loop},
  year          = {2026},
  howpublished  = {arXiv:2606.04997},
  eprint        = {2606.04997},
  archivePrefix = {arXiv},
  primaryClass  = {physics.atom-ph},
  doi           = {10.48550/arXiv.2606.04997}
}

@misc{Subramanian2026BMF,
  author        = {Subramanian, Keerthan and Tanaka, Hiroki
                   and Herr, Simon J. and Sah, Nutan Kumari
                   and Jha, Gaurav and Zacherl, Florian
                   and others},
  title         = {Generation of continuous-wave laser light
                   at 148.4 nm using cavity-enhanced second
                   harmonic generation in {BaMgF$_4$}},
  year          = {2026},
  howpublished  = {arXiv:2606.25046},
  eprint        = {2606.25046},
  archivePrefix = {arXiv},
  primaryClass  = {physics.optics},
  doi           = {10.48550/arXiv.2606.25046}
}

@article{Pan2002SBO,
  author  = {Pan, Feng and Shen, Guangqiu and Wang, Ruji
             and Wang, Xiaoqing and Shen, Dezhong},
  title   = {Growth, characterization and nonlinear optical
             properties of {SrB$_4$O$_7$} crystals},
  journal = {Journal of Crystal Growth},
  volume  = {241},
  number  = {1-2},
  pages   = {108--114},
  year    = {2002},
  doi     = {10.1016/S0022-0248(02)00873-4},
  publisher={Elsevier}
}

@misc{Vasilyev2026VUV,
  author       = {Sergey Vasilyev and Tian Ooi and Igor Moskalev and Mike Mirov
                  and Andrey Muraviev and Dmitrii Konnov and Victor Churikov
                  and Viktor Sukharev and Evgeny Galenin and Jack F. Doyle
                  and Chuankun Zhang and Kai Li and Georgiy Seryogin and Dan Perlov
                  and Igor Samartsev and Konstantin Vodopyanov and Jun Ye},
  title        = {Record nonlinear conversion efficiency in the production of
                  high spectral purity vacuum ultraviolet laser at 148 nm},
  howpublished = {arXiv:2606.19484},
  year         = {2026},
  doi          = {10.48550/arXiv.2606.19484}
}

@misc{YeICAP29_2026,
  author       = {Jun Ye},
  title        = {Laser-driven {M}{\"o}ssbauer spectroscopy for nuclear clock},
  howpublished = {Invited presentation at the 29th International Conference on Atomic Physics (ICAP29)},
  year         = {2026},
  month        = {June},
  note         = {Wuhan, China}
}

@misc{PerlovPatent2024,
  author = {Dan Perlov and Alexander Zaytsev and Anatolii Zamkov
            and Nikita Radionov and Aleksandr Cherepakhin
            and Nikolay Evtikhiev and Andrey Sadovskiy},
  title = {Method for manufacturing of patterned {SrB$_4$O$_7$} and {PbB$_4$O$_7$} crystals},
  year   = {2024},
  note   = {U.S. Patent 11,868,022}
}

\clearpage

\end{document}